\documentclass[aps,preprint,showpacs,preprintnumbers,amsmath,amssymb,showkeys,superscriptaddress,titlepage,secnumarabic]{revtex4-2}
\usepackage{graphicx}
\usepackage{dcolumn}
\usepackage{bm}
\usepackage{textcomp}
\usepackage{makecell}
\usepackage{hyperref}

\begin{document}
\title{Threshold resonances in relativistic \texorpdfstring{$^{16}$O}{16O} and \texorpdfstring{$^{22}$Ne}{22Ne} dissociation}

\author{D.A.~Artemenkov}
\affiliation{Joint Institute for Nuclear Research, Dubna Russia}

\author{N.K.~Kornegrutsa}
\affiliation{Joint Institute for Nuclear Research, Dubna Russia}

\author{V.V.~Rusakova}
\affiliation{Joint Institute for Nuclear Research, Dubna Russia}

\author{A.A.~Zaitsev\footnote{Corresponding author: zaicev@jinr.ru}}
\affiliation{Joint Institute for Nuclear Research, Dubna Russia}
\affiliation{Lebedev Physical Institute of the Russian Academy of Sciences, Moscow Russia}

\author{P.I.~Zarubin}
\affiliation{Joint Institute for Nuclear Research, Dubna Russia}
\affiliation{Lebedev Physical Institute of the Russian Academy of Sciences, Moscow Russia}

\begin{abstract}
The nuclear emulsion method is applied to study binary dissociation leading to the formation of $\alpha$-particles from $^{16}$O nuclei at 3.65 GeV per nucleon and $^{22}$Ne nuclei at 3.22 GeV per nucleon. The invariant mass distribution for the $^{16}$O~$\to$~$^{12}$C(0$^+_1$)+$\alpha$ channel indicates decays of the $^{16}$O(0$^+_6$) excitation, whose probability exceeds the estimate for the $^{12}$C(0$^+_2$)+$\alpha$ channel by a factor of 20. The invariant mass distribution for $^{22}$Ne~$\to$~$^{18}$O+$\alpha$ exhibits a peak corresponding to $^{22}$Ne excitations around 11.6 MeV.
\end{abstract}
\maketitle

\section{Introduction}
The use of nuclear emulsions (NE) with relativistic nuclear beams expands the study of nuclear clustering and, above all, of resonance states near the $\alpha$-particle separation thresholds. A recent review presents the genesis, current motivation, and status of research conducted with this approach \cite{nsr-2026}. In the BECQUEREL experiment, NE layers exposed from the early 1970s to the present with a wide variety of nuclei, including radioactive ones, are available for analysis, along with measurements performed on them. The identification of decays of the unstable nuclei $^{8}$Be, $^{9}$B, and the Hoyle state $^{12}$C(0$^+_2$) using the invariant mass method on this material \cite{Artemenkov-2020} indicates the completeness of the observations and the feasibility of identifying pairs involving heavier fragments below the Coulomb barrier. 

The current focus is on the search for extremely low-energy $\alpha$-particle states, described as Bose-Einstein condensates \cite{Tohsaki-2017}. Following the 2$\alpha$-state of $^{8}$Be(0$^+$) and the 3$\alpha$-state of Hoyle $^{12}$C(0$^+_2$), the excitation of $^{16}$O(0$^+_6$) at 15.1 MeV or 296 keV above the $^{12}$C(0$^+_2$)+$\alpha$ threshold has been proposed as a 4$\alpha$-condensate. 
The problem of the 5$\alpha$-condensate has been posed \cite{Zhou-2023}. These searches form part of the broader question of the existence of nuclear structures at densities significantly lower than those in the ground states of nuclei.

In NE layers longitudinally exposed to relativistic nuclei, the directions of generated fragments can be measured
with unique resolution and completeness. Figure 1 shows a macrophotograph of the splitting of a $^{16}$O nucleus at 3.65 GeV per nucleon into a pair of relativistic fragments C and He accompanied by the production of a meson pair ($s$-particles) and a recoil proton ($g$-particle). The tracks of relativistic He and H isotopes are identified visually, while for heavier nuclei, charge identification is possible via ionization measurements. Searching for interactions by sequentially tracking primary beam tracks allows the extraction of charge-channel fragmentation distributions, without sampling biases, which reflects the structure of the nuclei under study. The influence of dynamics is minimized in coherent dissociation events that contain no tracks of target fragments or produced mesons - so-called ``white'' stars. However, the fraction of stars in which the primary charge is transferred to fragments does not exceed a few percent, and the final states of interest are even rarer. The detection of relevant track groups can be accelerated by scanning the layers in a direction perpendicular to the incoming beam.

\begin{figure}
 \centering
        \includegraphics[width=0.7\textwidth]{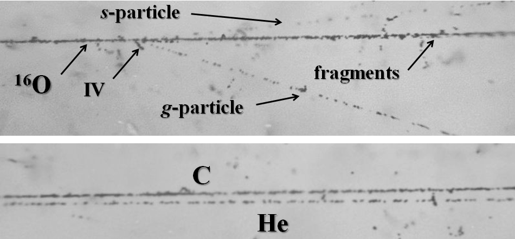}
 \caption{Macrophotograph of the splitting of a $^{16}$O nucleus at 3.65 GeV per nucleon into a pair of relativistic C and He fragments with production of a  meson pair ($s$-particles) and a recoil proton ($g$-particle). The interaction vertex (IV) and the resulting tracks are indicated, with the exception of one $s$-particle, whose track moves out of focus. As one proceeds along the C and He tracks, their separation is observed in the coordinate measurements up to 1 mm (bottom photo).}
\label{fig1}
\end{figure}

In general, decays of unstable states can be identified in distributions over the invariant mass $Q$, defined via sums of the scalar products of fragment $4$-momenta. For convenience, the masses of the fragments or the parent nucleus are subtracted from $Q$. In the relativistic case, the $4$-momenta are determined with sufficient accuracy from the emission angles under the assumption of conservation of momentum per nucleon of the parent nucleus. In the dissociation of $^{12}$C and $^{16}$O nuclei, the assumption that He fragments correspond to $\alpha$ particles is justified. Owing to their minimal energy, their decays can manifest themselves as ensembles of relativistic He fragments with the smallest opening angles.

Thus, the contribution of the $^8$Be(0$^+$)+$\alpha$ channel at $Q_{2\alpha}$~$\leq$~200 keV to the dissociation $^{12}$C~$\to$~3$\alpha$ at 3.65 GeV per nucleon was determined to be 43\%, including 26\% proceeding through $^{12}$C(0$^+_2$) at $Q_{3\alpha}$~$\leq$~700 keV. The contribution of $^{12}$C(0$^+_2$) to the coherent dissociation of $^{16}$O~$\to$~4$\alpha$ at 3.65 GeV per nucleon was 22\%, noticeably enhanced compared to the case of $^{12}$C (11\%). These findings suggest the generation of the unstable states of $^8$Be(0$^+$), $^{12}$C(0$^+_2$), and $^{16}$O(0$^+_6$) via sequential addition of the emerging $\alpha$-particles with decay in the reverse order. Alongside the near-threshold 4$\alpha$-decay of $^{16}$O(0$^+_6$), the decay  channels $^{12}$C(0$^+_1$)+$\alpha$, $^{12}$C(2$^+_1$)+$\alpha$ and $^{15}$N+$p$ with thresholds of 7.16, 11.60 and 12.13 MeV, respectively, cannot be ruled out.

However, $^{16}$O(0$^+_6$) at $Q_{4\alpha}$~$\leq$~1 MeV was not observed in the $^{12}$C(0$^+_2$)+$\alpha$ and $^8$Be(0$^+$)+$^8$Be(0$^+$) channels \cite{Artemenkov-2020}. This conclusion is consistent with the findings of low-energy experiments (e.g. \cite{Smith-2021}). Such suppression can be explained by the extremely small energy difference between $^{16}$O(0$^+_6$) and $^{12}$C(0$^+_2$)+$\alpha$, leading to a decay width of only 10$^{-9}$ MeV \cite{Funaki-2009}. At the same time, the structure of the $^{16}$O(0$^+_6$) state, proposed in the container model as $^{12}$C(0$^+_2$)+$\alpha$, allows the internal conversion of the $^{12}$C(0$^+_2$) sub-unit to the $^{12}$C(0$^+_1$) ground state with a width of 0.2 MeV \cite{Funaki-2009}. By allowing the $^{16}$O(0$^+_6$) state to be considered as an excitation near the 4$\alpha$ threshold, this scenario motivated the search for binary decays $^{16}$O(0$^+_6$)~$\to$~$^{12}$C(0$^+_1$)+$\alpha$ presented here.

The study of the binary $^{16}$O dissociation prompted a re-analysis of existing data on the $\alpha$-particle dissociation of the $^{22}$Ne nucleus \cite{El-Naghy-1988} for observations of near-threshold resonances, as follows. In this case, the $^{18}$O-$\alpha$ states near the coupling threshold are of interest, for which it is possible to determine the invariant masses $Q_{^{18}\mathrm{O}\alpha}$.

\section{Dissociation of the $^{16}$O nucleus}

Exposure of NE layers to $^{16}$O nuclei was performed in the 1980s at 3.65 GeV per nucleon at the JINR Synchrophasotron. An unbiased sample of 2823 stars included 21 ``white'' stars with a C~+~He topology and 9 with a 4$\times$He topology \cite{Andreeva-2005}. In the 1990s, the statistics of 4$\times$He ``white'' stars were increased to 641 via transverse scanning \cite{Andreeva-1996}. When revisiting these data in 2018, a significant contribution of the cascade channel $^{12}$C(0$^+_2$)+$\alpha$, mentioned above, was discovered in the invariant mass distribution of $\alpha$-triples. This finding was based on the identification of $^8$Be and $^9$B decays in the dissociation of $^9$Be, $^{10}$B, $^{10}$C and $^{11}$C in the BECQUEREL experiment. However, only 9 events with $Q_{4\alpha}$~$<$~1 MeV in the invariant mass distribution of $\alpha$-quartets could correspond to $^{16}$O(0$^+_6$)~$\to$~$^{12}$C(0$^+_2$)+$\alpha$. Despite the almost negative result for $^{16}$O(0$^+_6$), these data allow further comparison with the $^{12}$C(0$^+_1$)+$\alpha$ channel.

\begin{figure}
 \centering
        \includegraphics[width=0.8\textwidth]{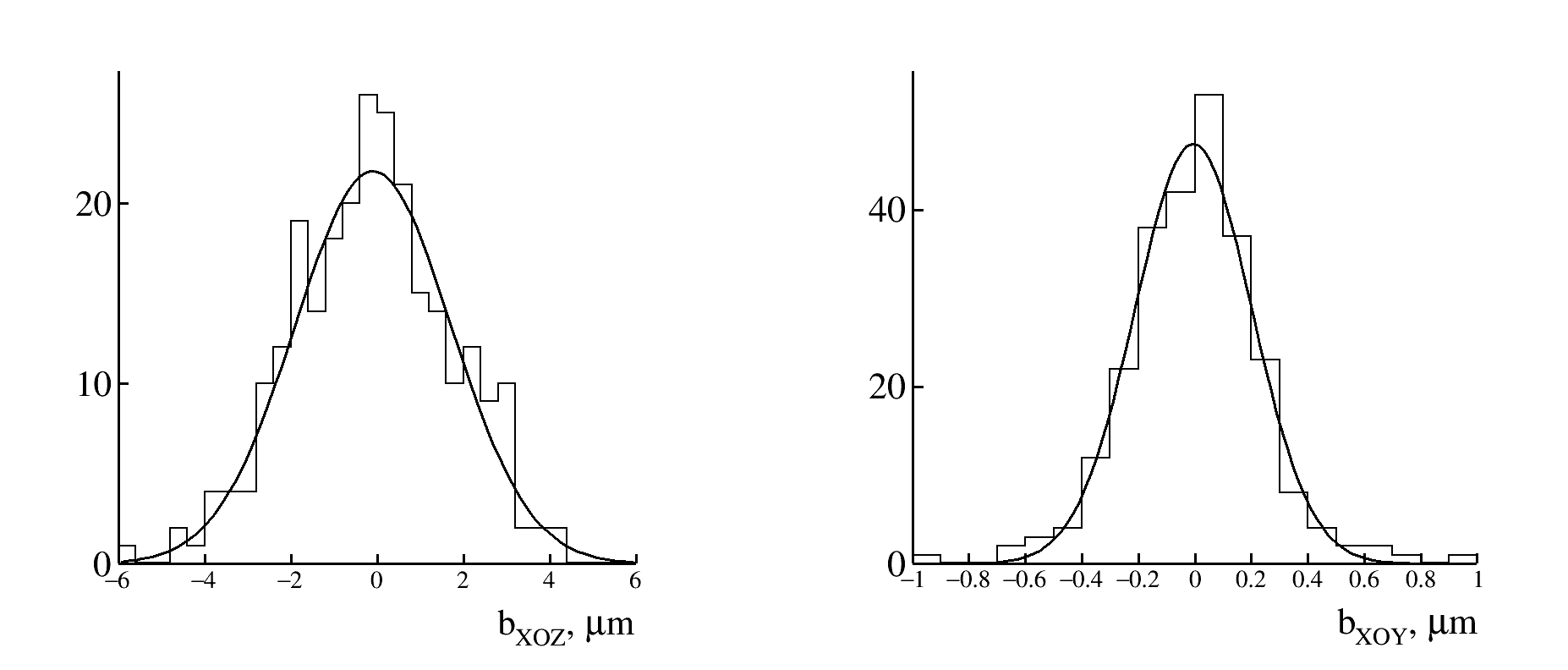}
 \caption{Distribution of distances $b_\mathrm{XOZ}$ over the layer depths and in their planes $b_\mathrm{XOY}$ between the observed positions of the vertices of the selected C~+~He pairs and their position reconstructed from the measured coordinates on the tracks.}
\label{fig2}
\end{figure}

In the present study, the statistics of C~+~He candidates were extended by cross-scanning to 200 pairs (examples in Fig. 1). The complexity of their interpretation as $^{12}$C~+~$\alpha$ was created by the overlap of 4He tracks associated with $^{16}$O. This problem was absent in the 4$\alpha$ case, due to the clear definition of the vertices and the ionization difference. The selection was based on the differences $b_\mathrm{XOZ}$ and $b_\mathrm{XOY}$ (Fig. 2) between the vertex coordinates measured on the KSM microscope and their positions determined by straight lines approximating the coordinates on the tracks with a step of 100 $\mu$m over lengths from 1 to 4 mm. In the planar XOY plane, the discrepancy in $b_\mathrm{XOY}$ is comparable to the NE grain size. In the perpendicular XOZ plane, $b_\mathrm{XOZ}$ is an order of magnitude larger due to the shrinkage of the developed layers, focusing noise, and vibrations of the microscope stage. These factors lead to an interpolation error of several microns. For the overwhelming majority of selected events, the accuracy of vertex position determination was 0.6~$\pm$~0.3 $\mu$m in the XOY plane and 5.1~$\pm$~0.3 $\mu$m in the XOZ plane. 

The separation of charges on the primary and secondary tracks, performed by visual counting the density of $\delta$-electrons $N_{\delta}$(O)/100 $\mu$m, allowed us to strengthen the selection for 46 stars (Fig. 3). For beam particles, the average number $\langle N_{\delta}(\mathrm{O})/100\rangle$ is 4.8~$\pm$~0.1 with the parameter of the fitted Gaussian distribution $\sigma$~=~0.6~$\pm$~0.1, and for fragments $\langle N_{\delta}(\mathrm{O})/100\rangle$~=~2.2~$\pm$~0.1 with $\sigma$~=~0.5~$\pm$~0.1.

The planar angles $\alpha$ and the immersion angles $\phi$ relative to the primary tracks were calculated using a linear approximation of the coordinates on the tracks of 130 selected pairs, including 42 ``white'' stars. From these, the polar emission angles $\theta_\mathrm{C}$ and $\theta_\mathrm{He}$ were derived (Fig. 4). The distributions over $\theta_\mathrm{C}$ and $\theta_\mathrm{He}$ are described by a Rayleigh distribution with parameters $\sigma_{\theta_\mathrm{C}}$~=~(4.1~$\pm$~0.2)$\times$10$^{-3}$ rad and $\sigma_{\theta_\mathrm{He}}$~=~(8.3~$\pm$~0.7)$\times$10$^{-3}$ rad. The distributions over the spatial $\Theta_\mathrm{CHe}$ and azimuth angle $\Psi_\mathrm{CHe}$ between fragments C and He are shown in Figs. 5 and 6, respectively. In the region of the peak discussed below at $Q_{\alpha^{12}\mathrm{C}}$~$\approx$~15 MeV, the average value of $\langle \Theta_{\mathrm{CHe}} \rangle$ is equal to (15.5~$\pm$~0.4)$\times$10$^{-3}$ rad, with the parameter of the approximating Gaussian distribution $\sigma$~=~(1.5~$\pm$~0.4)$\times$10$^{-3}$ rad.

\begin{figure}
 \centering
        \includegraphics[width=0.5\textwidth]{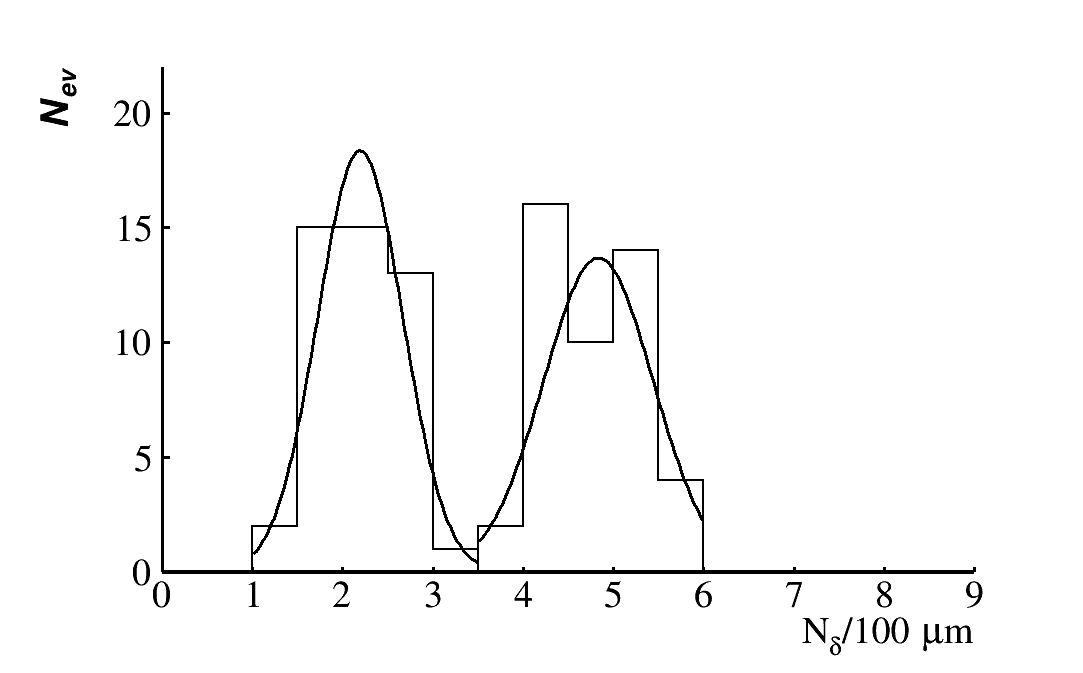}
 \caption{Distribution over $\langle N_{\delta}/100\rangle$ for primary tracks and tracks of secondary heavy fragments in the dissociation $^{16}$O~$\to$~C + He; the curves are Gaussian fits.}
\label{fig3}
\end{figure}

\begin{figure}
 \centering
        \includegraphics[width=0.5\textwidth]{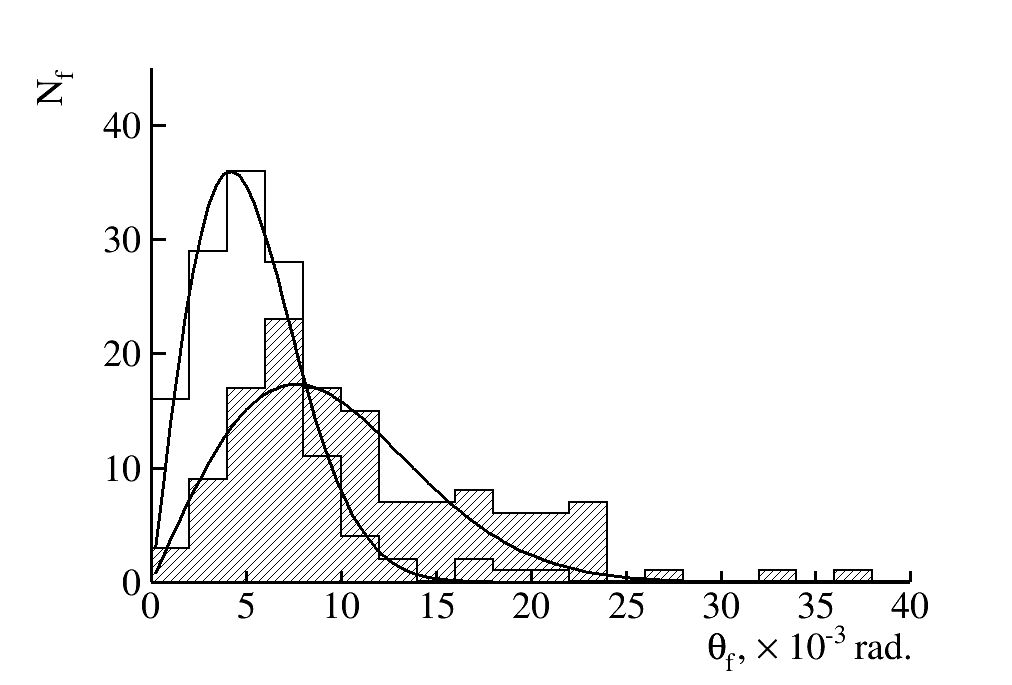}
 \caption{Distributions of fragments over polar angles $\theta_\mathrm{C}$ and $\theta_\mathrm{He}$ (shaded); the curves are Rayleigh fits.}
\label{fig4}
\end{figure}

\begin{figure}
 \centering
        \includegraphics[width=0.5\textwidth]{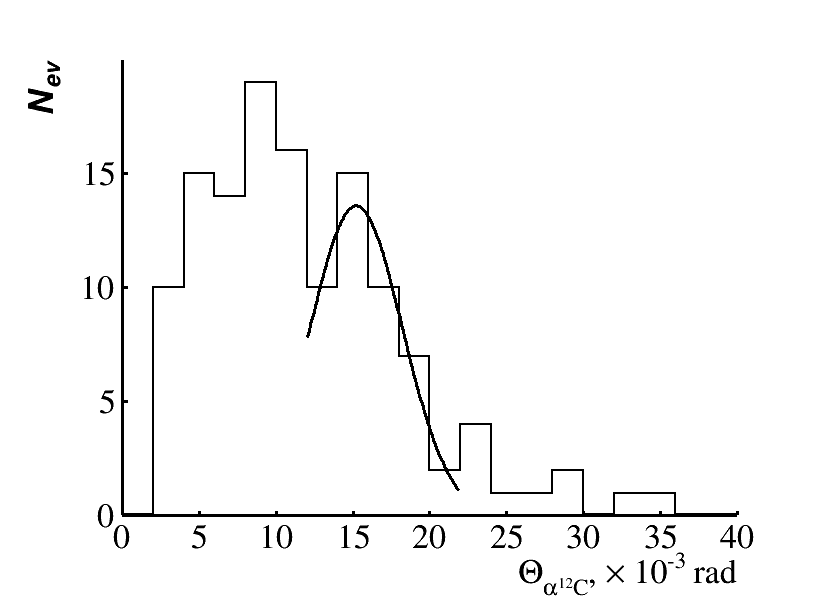}
 \caption{Distribution over the spatial angle $\Theta_\mathrm{CHe}$ between C and He fragments; the curve is a Rayleigh fit in the region of the peak at $Q_\mathrm{\alpha^{12}C}$~$\approx$~15 MeV.}
\label{fig5}
\end{figure}

\begin{figure}
 \centering
        \includegraphics[width=0.5\textwidth]{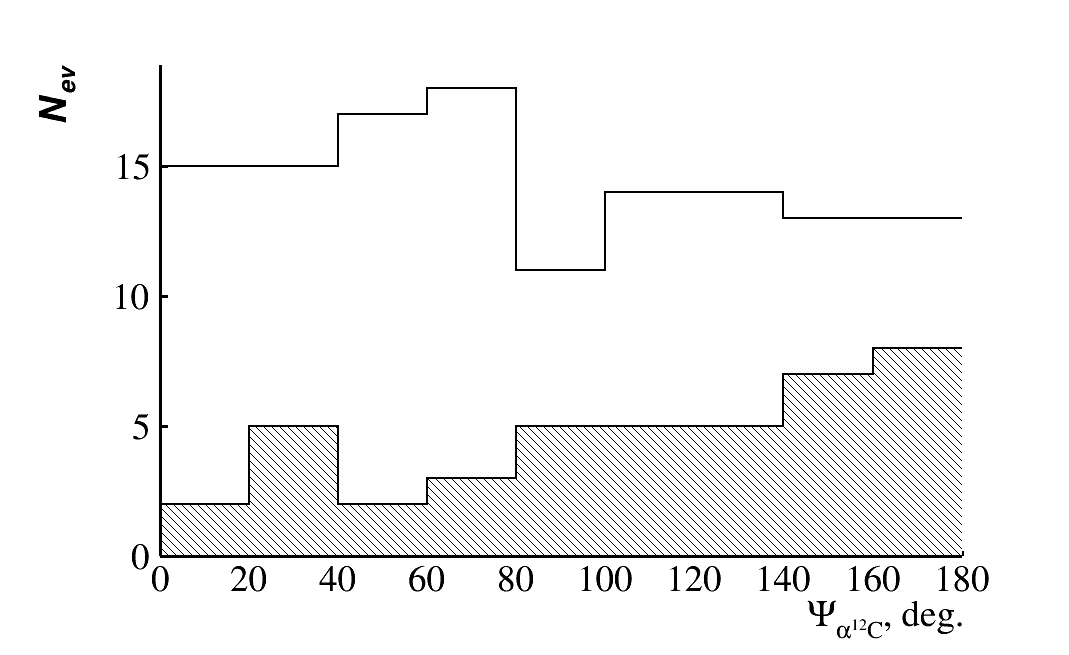}
 \caption{Distribution of C~+~He pairs over the azimuthal angle $\mathrm\psi$; ``white'' stars are highlighted with hatching.}
\label{fig6}
\end{figure}

The C~+~He channels with neutron separation have thresholds $^{11}$C + $^4$He (23.5 MeV) and $^{12}$C + $^{3}$He (27.8 MeV), which significantly exceed that of $^{12}$C + $\alpha$ (7.16 MeV). According to the isotopic identification of $^{16}$O~$\to$~C + He at 2.5 GeV per nucleon performed in a hydrogen bubble chamber in a magnetic field \cite{Glagolev-2021}, the role of $^{11}$C and $^{3}$He is negligible \cite{nsr-2026}. Therefore, assuming that the C fragments correspond to the isotope $^{12}$C and He fragments to $\alpha$-particles, Fig. 7 shows the distribution of the pairs over the invariant mass $Q_{^{12}\mathrm{C}\alpha}$. The $Q_{^{12}\mathrm{C}\alpha}$ values were calculated from the emission angles of $^{12}$C and $\alpha$-particles in the approximation of conservation of momentum per nucleon of the $^{16}$O nucleus. In the range 13~$< Q_{^{12}\mathrm{C}\alpha} <$ 17 MeV, a peak is observed with an average value of $\langle Q_{^{12}\mathrm{C}\alpha}\rangle$~= 14.9~$\pm$~0.4 MeV. The parameter $\sigma$ of the fitted Gaussian distribution is 1.2~$\pm$~0.4 MeV. Nineteen $^{12}$C+$\alpha$  pairs in the peak can be attributed to decays of $^{16}$O(0$^+_6$)~$\to$~$^{12}$C(0$^+_1$)+$\alpha$, including 7 ``white'' stars. Proportional recalculation of the statistics of ``white'' stars here and in the above-mentioned sequential and transverse searches, taking into account the difference in selection in them, gives a ratio of the probabilities of the channels $^{16}$O(0$^+_6$)~$\to$~$^{12}$C(0$^+_1$)+$\alpha$ and $^{16}$O(0$^+_6$)~$\to$~$^{12}$C(0$^+_2$)+$\alpha$ of at least 20.

Figure 8 shows the distribution over the total transverse momentum $P_{T(^{12}\mathrm{C}+\alpha)}$ pairs. It is described by a Rayleigh distribution with a parameter $\sigma$~=~289~$\pm$~2 MeV/$c$. This value is characteristic of nuclear diffraction interactions, determined by the NE composition. However, in the case of ``white'' stars, the proportion of pairs with $P_{T(^{12}\mathrm{C}+\alpha)} <$~150 MeV/$c$, which can arise in electromagnetic dissociation, increases.

\begin{figure}
 \centering
        \includegraphics[width=0.65\textwidth]{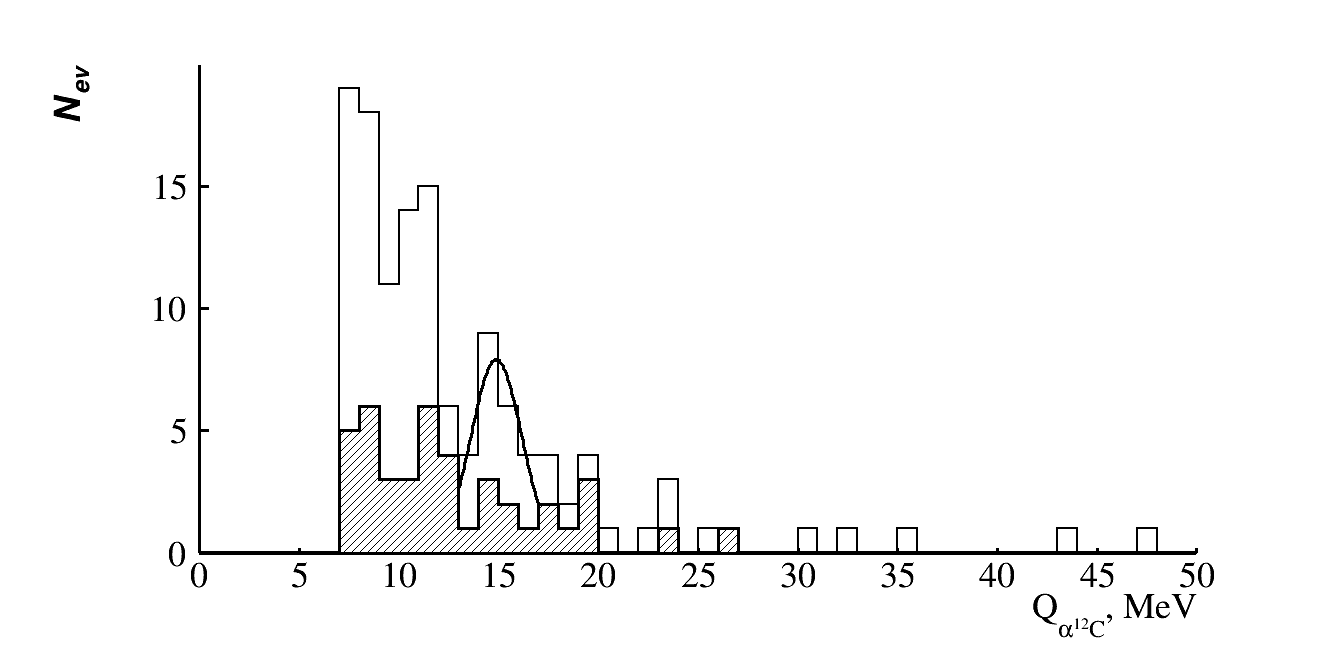}
 \caption{Distribution of C~+~He pairs in the dissociation of $^{16}$O nuclei at 3.65 GeV per nucleon over the invariant mass $Q_{\alpha^{12}\mathrm{C}}$; the contribution of ``white'' stars is hatched.}
\label{fig7}
\end{figure}

\begin{figure}
 \centering
        \includegraphics[width=0.5\textwidth]{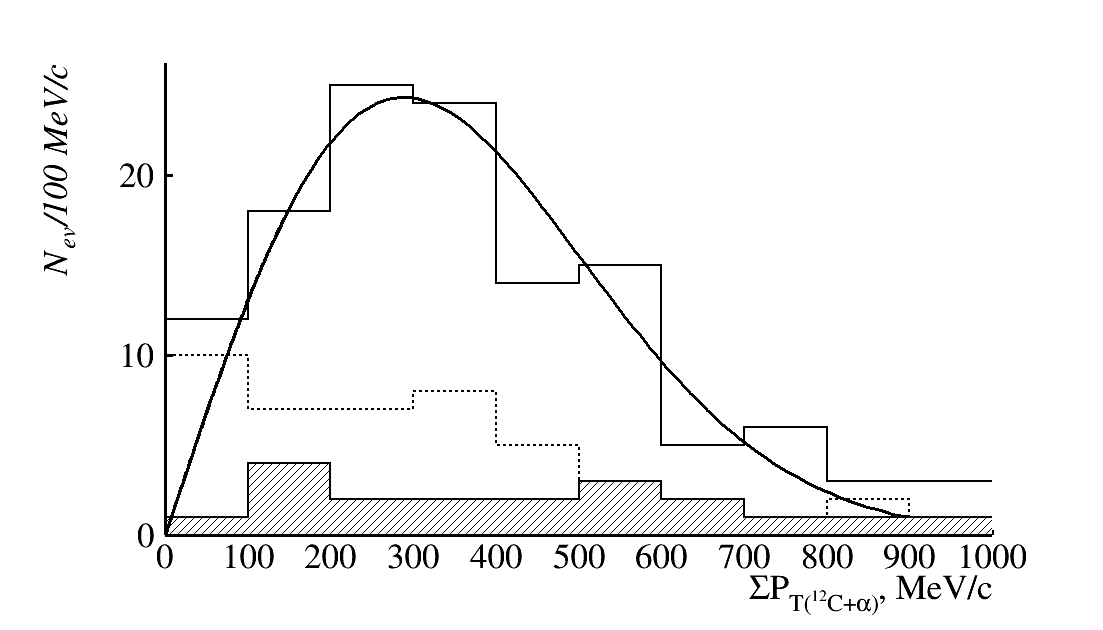}
 \caption{Distribution of C~+~He pairs in the dissociation of $^{16}$O nuclei at 3.65 GeV per nucleon over the total transverse momenta $P_{T(^{12}\mathrm{C}+\alpha)}$ (solid line), including ``white'' stars (dashed line) or events 13~$ < Q_{^{12}\mathrm{C}\alpha} <$~17 MeV (hatched line).}
\label{fig8}
\end{figure}

The $Q_{\alpha^{12}\mathrm{C}}$ distribution contains a group of pairs that can be attributed to decays of short-lived states from the first $\alpha$-unbound level at 8.87 MeV to the $^{15}$N+$p$ threshold. Such pairs could arise from the final state interactions between $^{12}$C and $\alpha$. The presence of pairs with lower $Q_{^{12}\mathrm{C}\alpha}$ values is unexpected. According to hydrogen bubble chamber data, they do not appear in dissociation on protons \cite{nsr-2026}. A possible explanation is bremsstrahlung of some $^{12}$C fragments in the field of target nuclei. This effect may not be evident in $^{16}$O spectroscopy, but manifests itself in the calculation of the invariant masses of the final states.

\section{Dissociation of the $^{22}\mathrm{\textbf{Ne}}$ nucleus}

The preserved measurements of 4307 stars produced in NE by $^{22}$Ne nuclei at 3.22 GeV per nucleon \cite{El-Naghy-1988} allow us to continue the search for near-threshold resonances in heavy fragment and $\alpha$-particle pairs. Among them, the charge 10 was transferred to relativistic fragments in 855 stars, including 121 ``white'' stars (17\%). The leading ones are 158 pairs O~+~He, including 54 ``white'' stars (34\%) \cite{El-Naghy-1988,Andreeva-2005}. Initially, this feature was attributed to splitting into a pair of doubly magic nuclei $^{16}$O~+~$^4$He~+~(2$n$). However, the breakup threshold of $^{22}$Ne into $^{18}$O$\alpha$, equal to 9.7 MeV, is noticeably lower than those for $^{16}$O$\alpha$2$n$ (16.5), $^{17}$O$\alpha n$ (17.7 MeV) and $^{16}$O$^6$He (20.9 MeV). Its value indicates a strong binding of the neutron pair to the $^{16}$O core and the possibility of the formation of $^{18}$O$\alpha$ resonance.

The lack of the $^{18}$O isotope identification was proposed to be compensated by selecting O and He fragments with relatively opposite directions in the azimuthal angle $\Delta\mathrm\psi_{\mathrm{O+He}}$ to enhance the binary splitting signal (Fig. 9). Of significant importance is the veto in the O~+~He channel of the formation of charged mesons $n_s$ = 0 ($s$-particles) and target fragments $n_h$ = 0 ($h$-particles), which could otherwise be affected by neutron contributions. Indeed, the $\Delta\psi_{\mathrm{O+He}}$ distribution exhibits an asymmetry under the combined condition $n_s$ = 0 and $n_h$ = 0.

\begin{figure}
 \centering
        \includegraphics[width=0.5\textwidth]{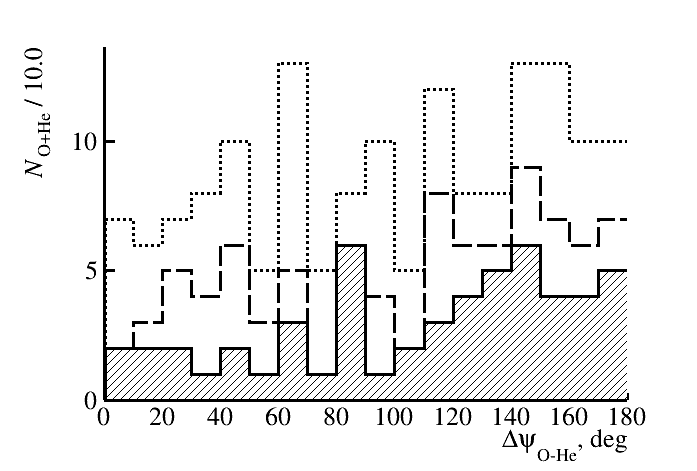}
 \caption{Distribution of O~+~He pairs over the azimuthal angle $\Delta\psi_{\mathrm{O+He}}$ (dots), including $n_s$ = 0 (dashed line) and ``white'' stars (solid line with hatching).}
\label{fig9}
\end{figure}

\begin{figure}
 \centering
        \includegraphics[width=0.6\textwidth]{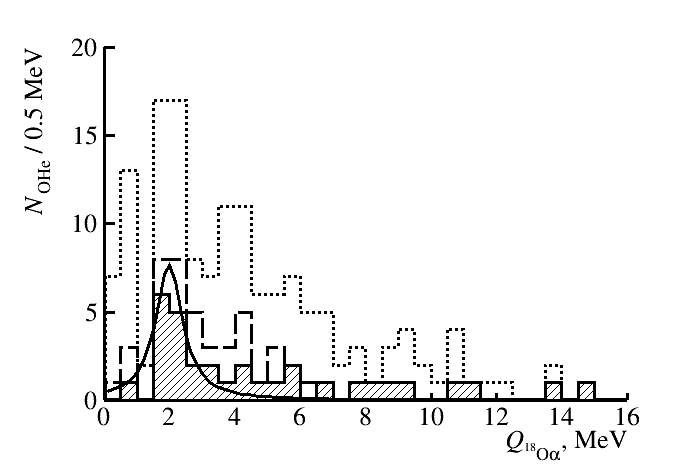}
 \caption{Distribution of O~+~He pairs from the dissociation of $^{22}$Ne nuclei at 3.22 GeV per nucleon over the invariant mass $Q_{^{18}\mathrm{O}\alpha}$ (dots), ``white'' stars (dashed line) and ``white'' stars (hatched line) with the condition $\Delta\psi_{\mathrm{O+He}} > $ 100$^\circ$.}
\label{fig10}
\end{figure}

Figure 10 shows the distribution of all O~+~He with $\Delta\psi_{\mathrm{O+He}} > $ 100$^\circ$ pairs over the invariant mass $Q_{^{18}\mathrm{O}\alpha}$ calculated under the approximation of conservation of the initial momentum per nucleon and the assumption of $^{18}$O$\alpha$ correspondence. The selection of ``white'' stars, corresponding to the minimal excitation of $^{22}$Ne, practically eliminates the broad maximum at large values of $Q_{^{18}\mathrm{O}\alpha}$ and retains the narrow maximum at lower values. The condition $\Delta\psi_{\mathrm{O+He}} > $ 100$^\circ$ did not lead to the disappearance of the latter. Given a possible overestimation of the isotope mass number as $^{18}$O, $Q_{^{18}\mathrm{O}\alpha}$ serves as a minimum estimate of $Q$ since taking into account neutrons in the invariant mass would increase it.

A fit of the narrow peak with the Breit-Wigner function, with the fixed width at $\Gamma$~=~1.0 MeV, gives the parameter $\langle Q_{^{18}\mathrm{O}\alpha} \rangle$~=~1.97~$\pm$~0.15 MeV (Fig. 10), providing grounds to consider the pairs in this peak as the decay product of one or several close states near the threshold. The distribution of the pairs over the total transverse momentum $P_{T(^{18}\mathrm{O}+\alpha)}$ is shown in Fig. 11. It is described by the sum of two Rayleigh functions with parameters $\sigma$~=~67~$\pm$~23 and 260~$\pm$~49 MeV/$c$, characteristic for electromagnetic and nuclear diffraction dissociation, respectively. The ratio of these mechanisms is determined by the NE composition.

\begin{figure}
 \centering
        \includegraphics[width=0.55\textwidth]{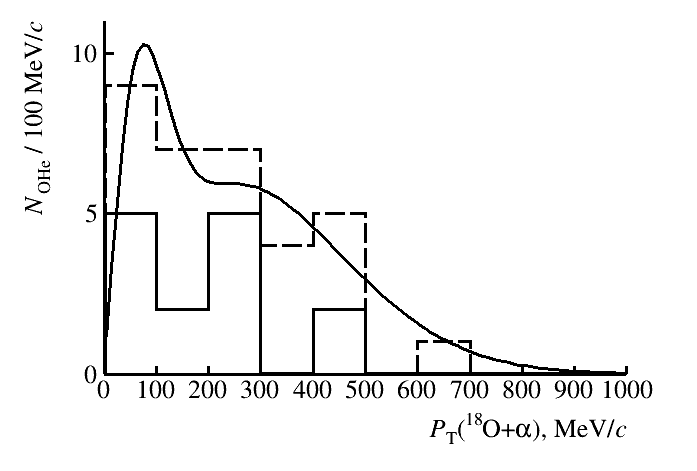}
 \caption{Distribution of ``white'' O~+~He stars in the dissociation of $^{22}$Ne nuclei at 3.22 GeV per nucleon over the total transverse momentum $P_{T(^{18}\mathrm{O}+\alpha)}$ (dashed line), including events $Q_{\alpha^{18}\mathrm{O}} < $~3 MeV (solid line); the curve is a fit using two Rayleigh functions.}
\label{fig11}
\end{figure}

\section{Conclusions}

Previously, our NE based studies of the unstable states focused on the lightest isotope ensembles. The minimum threshold values for the separation of $\alpha$-particles allow this approach to be extended to the binary $^{16}$O and $^{22}$Ne dissociation bypassing the residue isotopic identification.

The invariant mass distribution of $^{12}$C+$\alpha$ pairs produced in the dissociation of $^{16}$O at 3.65 GeV per nucleon provides evidence for decays of the excited state $^{16}$O(0$^+_6$). An estimate of the probabilities of the $^{12}$C(0$^+_1$)+$\alpha$ and $^{12}$C(0$^+_2$)+$\alpha$ channels indicates a dominance of the former by a factor of about 20. Despite statistical limitations, this conclusion motivates further $^{12}$C(0$^+_1$)+$\alpha$ studies at lower incident energies. It supports the hypothesis of decay of the $^{16}$O(0$^+_6$) state via internal conversion of its $^{12}$C(0$^+_2$) substructure to $^{12}$C(0$^+_1$). This opens up the possibility of synthesizing the isotope $^{12}$C through the decay $^{16}$O(0$^+_6$)~$\to$~$^{12}$C(0$^+_2$)+$\alpha$~$\to$~$^{12}$C(0$^+_1$)+$\alpha$ without a radiative transition, as a supplement to the known 3$\alpha$-process.

The invariant mass distribution of $^{18}$O$\alpha$ pairs produced in the dissociation of $^{22}$Ne at 3.22 GeV per nucleon exhibits a peak that can be interpreted as $^{22}$Ne excitations with a mean value of about 11.6 MeV. This energy corresponds to the sub-barrier \(^{18}\mathrm O+\alpha\) fusion region and provides further evidence for the feasibility of nucleosynthesis reactions in the fragmentation cone of relativistic nuclei.

While not serving as an alternative to low-energy experiments, the approach presented here allows us to complement the pattern of relativistic nuclear collisions with insights into cluster excitations that are important in nuclear astrophysics. Recent studies of $^{22}$Ne near-threshold excitations \cite{Nurmukhanbetova-2024,Greaves-2026} suggest that the search for near-threshold resonances deserves to be continued in the dissociation of $^{18}$O~$\to$~$^{14}$C~+~$\alpha$ and $^{14}$C~$\to$~$^{10}$Be~+~$\alpha$. In these cases, the analysis will be simpler than for the $^{22}$Ne case, but new NE exposures are required. 

It should be noted that the choice of several GeV per nucleon in the context of nuclear clustering is determined by historical reasons. The invariant mass distribution of $^{12}$C~$\to$~3$\alpha$ dissociation events in nuclear emulsion at 0.45 and 3.65 GeV per nucleon indicates a coincidence of the contributions from $^{12}$C(0$^+_2$) and $^{12}$C(3$^-$) decays  \cite{nsr-2026}. Longitudinal NE exposures should be performed in the region of several hundred MeV per nucleon where achieving beam profile uniformity and isotope separation is easier. Event-by-event correction of the beam nucleus energy for deceleration in NE is possible when calculating invariant masses.

\end{document}